\documentclass[prologue,table,xcdraw,sigconf]{acmart}

\usepackage{balance} 
\usepackage{xcolor}
\usepackage{multicol}
\usepackage{afterpage}

\usepackage{fancyhdr}
\usepackage{pdfcomment}
\usepackage{graphicx} 
\usepackage{geometry}
\usepackage{multirow}
\usepackage{subcaption}
\usepackage{hyperref}
\newcommand{\secref}[1]{\S\ref{#1}}
\usepackage{caption}
\usepackage{subcaption}
\usepackage{accsupp}

\usepackage[T1]{fontenc}
\usepackage{xparse}
\usepackage{enumitem}
\setlist[description]{
  font={\sffamily\bfseries},
  labelsep=25pt,
  labelwidth=\transcriptlen,
  leftmargin=!,
}

\newlength{\transcriptlen}

\NewDocumentCommand {\setspeaker} { mo } {%
  \IfNoValueTF{#2}
  {\expandafter\newcommand\csname#1\endcsname{\item[#1:]}}%
  {\expandafter\newcommand\csname#1\endcsname{\item[#2:]}}%
  \IfNoValueTF{#2}
  {\settowidth{\transcriptlen}{#1}}%
  {\settowidth{\transcriptlen}{#2}}%
}

\usepackage[normalem]{ulem}

\usepackage{pdfcomment}

\usepackage{makecell} 
\usepackage{array}    

\copyrightyear{2024}
\acmYear{2024}
\setcopyright{rightsretained}
\acmConference[ASSETS '24]{The 26th International ACM SIGACCESS Conference on Computers and Accessibility}{October 27--30, 2024}{St. John's, NL, Canada}
\acmBooktitle{The 26th International ACM SIGACCESS Conference on Computers and Accessibility (ASSETS '24), October 27--30, 2024, St. John's, NL, Canada}
\acmDOI{10.1145/3663548.3675633}
\acmISBN{979-8-4007-0677-6/24/10}

\begin{document}
\title{TwIPS: A Large Language Model Powered Texting Application to Simplify Conversational Nuances for Autistic Users}

\author{Rukhshan Haroon}
\affiliation{%
  \institution{Tufts University}
  \city{Medford}
  \state{Massachusetts}
  \country{USA}
  \postcode{02155}
}
\email{rukhshan.haroon@tufts.edu}

\author{Fahad Dogar}
\affiliation{%
  \institution{Tufts University}
  \city{Medford}
  \state{Massachusetts}
  \country{USA}
  \postcode{02155}
}
\email{fahad.dogar@tufts.edu}

\begin{abstract}

Autistic individuals often experience difficulties in conveying and interpreting emotional tone and non-literal nuances. Many also \textit{mask}\footnote{Masking entails consciously or unconsciously altering one's behavior to conform to societal expectations.}  their communication style to avoid being misconstrued by others, spending considerable time and mental effort in the process. To address these challenges in text-based communication, we present \textit{TwIPS}\footnote{\underline{T}exting \underline{w}ith \textsc{\underline{I}nterpret}, \textsc{\underline{P}review}, and \textsc{\underline{S}uggest}}, a prototype texting application powered by a large language model (LLM), which can assist users with: a) deciphering tone and meaning of incoming messages, b) ensuring the emotional tone of their message is in line with their intent, and c) coming up with alternate phrasing for messages that could be misconstrued and received negatively by others. We leverage an AI-based simulation and a conversational script to evaluate TwIPS with 8 autistic participants in an in-lab setting. Our findings show TwIPS enables a convenient way for participants to seek clarifications, provides a better alternative to tone indicators, and facilitates constructive reflection on writing technique and style. We also examine how autistic users utilize language for self-expression and interpretation in instant messaging, and gather feedback for enhancing our prototype. We conclude with a discussion around balancing user-autonomy with AI-mediation, establishing appropriate trust levels in AI systems, and autistic users' customization needs in the context of AI-assisted communication. 






\end{abstract}

\date{September 2023}

\maketitle

\section{Introduction}

Autism Spectrum Disorder (ASD) is a complex neuro-developmental disorder characterized by challenges in verbal and nonverbal communication, difficulties in social interactions, repetitive behaviors, and/or sensory sensitivities \cite{dsm5, grossi-patterns-of}. It is one of the most common neuro-developmental disorders in the United States, currently observed at a prevalence rate of approximately 1 in 36 among 8-year-old children \cite{cdc-prevalence}. 
Many autistic individuals find it challenging to process non-verbal cues, such as facial expressions and body language, in face-to-face (FTF) interactions. Variations in vocal pitch and tone can add more layers of complexity to communication, making FTF interactions overwhelming for them. Prior work in disabilities and linguistics underscores a preference for written communication among autistic individuals, highlighting a tendency towards email and text messaging over FTF interactions \cite{howard-anything-but}. This preference is attributed to the greater control and sensory ease provided by written communication \cite{nicolaidis-respect-healthcare}. However, studies indicate autistic individuals often experience difficulties in conveying and interpreting emotional tone and non-literal nuances in text-based communication, and standard chat features like GIFs and emojis contribute to these challenges instead of addressing them \cite{barros-my-perfect, page-perceiving-affordances}. Many also mask their writing style to avoid being misconstrued by others, spending considerable time and mental effort in the process.


In the past, technology has been extensively utilized to enhance autism diagnosis methods \cite{debelen-eyexplain-autism, arshad-east-early}, therapeutic interventions \cite{gesture-therapy, jeong-lexical-representation}, social support tools \cite{tartaro-virtual-peer, washington-a-wearable}, and the overall quality of life \cite{anxiety-autism, kim-routineaid-externalizing} of autistic individuals. Accessibility researchers have also advocated for the redesign of existing, mainstream applications to better meet the usability needs of autistic users \cite{grynszpan-human-computer, rapp-interactive-urban, kender-autism-design}. 
More recently, advances in large language models (LLMs) such as ChatGPT \cite{partha-chatgpt-a} have opened up new possibilities for AI-assisted assistive communication. Their applications range from helping dyslexic individuals write emails \cite{goodman-lampost-design} to providing social communication support for autistic individuals in professional settings \cite{jiwoong-its-the}. Thus, LLMs have demonstrated their capability to interpret and generate text with a level of nuance that rivals human abilities in various practical scenarios \cite{gpt4-openai-and}.
Despite these advances, text-based communication challenges for autistic users persist, making it crucial and timely to investigate how state of the art in LLMs can be used to address them.


Informed by specific communication challenges and design needs of autistic users identified in prior work \cite{barros-my-perfect, page-perceiving-affordances}, we introduce \textit{TwIPS}, a novel texting application that leverages recent advances in LLMs \cite{gpt4-openai-and} to grasp nuanced and implicit elements like emotional tone and intent from text messages. Based on its understanding of these elements, it can dynamically generate feedback for users that is tailored to the specificities of each conversation. TwIPS comprises of three features (described in detail in \secref{sec:prototype-overview}) not found in traditional texting applications, all of which are powered by GPT-4 \cite{gpt4-openai-and}:
\begin{enumerate}
    \item {\sc Interpret}: describes the overall tone and meaning of an incoming message, as well as individually identifies and explains ambiguous language elements in it, such as figurative language, sarcasm, and emojis;
    \item {\sc Preview}: allows users to preview the recipient's likely emotional reaction to their message, helping them verify whether the emotional tone of their message comes across to the recipient as intended;
    \item {\sc Suggest}: complements {\sc Preview} by suggesting a differently phrased alternate message when needed - the suggested message preserves the intent of the original message, but has a softer tone.
\end{enumerate}
As these features are in place, users retain full autonomy and agency to determine whether to adopt a suggestion or disagree with the provided feedback and disregard it entirely.

Through an in-lab user-study with 8 autistic participants recruited from a university setting, ranging in age from 18 to 44, we collected survey and in-depth qualitative data on a) their everyday use of language for self-expression and interpretation in instant messaging, and b) their perceptions of autonomy while using TwIPS, its usefulness, and suggestions for improving it. The user-study leveraged a conversational script and an AI-based simulation to present participants with scenarios where they could utilize each of TwIPS' three features, while maintaining user-autonomy, reproducibility and dynamicity across their experiences. Qualitative data showed participants demonstrated a strong inclination to actively engage in conversations instead of letting them become one-sided, though they found maintaining this balance difficult in text-based interactions. They also expressed their preference for clear and direct communication, emphasized the importance of punctuation in personalization and emotional expression, and shared their strategies for masking while texting. For many participants, TwIPS enabled a convenient way for them to seek clarifications, provided a better alternative to tone indicators, and facilitated constructive reflection on writing technique. Participants' suggestions for improving it centered around enhancing personalization and implementing measures to prevent users from over-relying on it. Our post-study 7-point Likert scale survey revealed 7 out of 8 participants favored continued use of TwIPS' all three features in their everyday chatting apps, with {\sc{Interpret}} receiving the highest rating (avg. 6.25, s.d = 1.03), followed by {\sc{Preview}} (avg. = 5.88, s.d = 1.64), and {\sc{Suggest}} (avg = 5.13, s.d = 1.72) \footnote{For each feature, participants were asked to express their level of disagreement (1) or agreement (7) on a Likert scale with the following statement, "I would like to continue using the \textit{X} feature in my everyday chatting app", where \textsc{X} was the name of the feature.}. 

Our work integrates concepts from both neurodiversity \cite{damian-on-the} and interventionist \cite{hogan-social-and} frameworks. TwIPS provides targeted assistance to users akin to an interventionist approach, while firmly upholding neurodiversity principles. This involves encouraging understanding others' perspectives, prioritizing user-autonomy and subjective feedback over rigid directives as a fundamental design choice, and recognizing communication as a collaborative process that requires mutual effort from all involved.

To summarize, we make the following key contributions: 
1) we design and implement TwIPS, an LLM-powered texting application, tailored to better meet the communication needs of autistic users, 2) we devise a novel methodology, leveraging an AI-based simulation and a conversational script, to evaluate all of TwIPS' features in an in-lab setting with 8 autistic participants, while maintaining user-autonomy, reproducibility and dynamicity across their experiences, 3) we conduct a thorough exploration into autistic individuals' use of language for self-expression and interpretation in instant messaging, closely examining the nuances of their communication styles, and 4) we gather in-depth feedback on user perceptions of autonomy and usefulness of AI-assistance in the context of instant messaging, identifying design and practical implications for augmenting text-based communication platforms with LLMs.

\section{Related Work}
In this section, we review the communication preferences of autistic individuals, ongoing efforts to integrate AI assistance into human writing, and broader initiatives leveraging computing technologies in the context of ASD.
\subsection{Communication Preferences in ASD}
Numerous studies in disabilities and linguistics research have explored the communication preferences of autistic individuals. Howard et al. conducted a study involving 245 autistic adult participants, and on average, these individuals selected email and text-messaging as their preferred modes of communication over FTF conversations \cite{howard-anything-but}. Nicolaidis et al. showed that there existed a link between the perceived success of healthcare interactions among autistic adults and the availability of written communication options \cite{nicolaidis-respect-healthcare}. Researchers posit that this is because written communication provides a higher degree of control, clarity, thinking time and sensory calm than FTF conversations \cite{benford-the-internet, gillespie-intersections-between}. As a result, autistic individuals often rely on text-based digital communication modes, such as email and instant messaging features supported by social networking platforms, for interacting with others \cite{burkey-social-use}. 

Given the popularity of text-based communication among autistic individuals, it becomes important to investigate whether the design of platforms that support text-based communication align with the needs and preferences of autistic individuals. A limited body of research in Human-Computer Interaction (HCI) literature attempts to answer this question through exploratory studies. Barros et al., in their critique of mainstream social media platforms, revealed that interpreting meaning and expressing emotional intent on social media platforms is particularly difficult for autistic users \cite{barros-my-perfect}. Page et al. reaffirmed these challenges, and emphasized that non-literal nuances, such as sarcasm and jokes, are even harder to comprehend \cite{page-perceiving-affordances}. It is well known that autistic individuals may exhibit a cognitive style that is characterized by a preference for literal thinking and a tendency to interpret information in a concrete and straightforward manner \cite{howard-anything-but}. Our research takes findings from these exploratory studies as a foundation to guide the design of an LLM-powered texting application tailored to cater to some of the challenges identified above. 




\subsection{AI Mediated Writing}


AI has long been applied to enhance digital written communication. Writing assistants have traditionally been used to predict short, one word suggestions, like the next probable word in a phrase \cite{philip-cost-benefit}. These are most common in emailing platforms, and Google has already deployed one on Gmail with access to millions of users \cite{henderson-efficient-natural}. The emergence of LLMs such as GPT-4 and BERT, however, is shifting this landscape \cite{gpt4-openai-and, jacob-bert-pretraining}. These models can not only produce longer pieces of text that are seemingly indistinguishable from human-written text, including the entire next likely sentence and creative content like poetry or story outlines, but also possess the capability to understand nuanced and implicit aspects such as intent and context. This is enabled
by massive, attention based transformer models that are trained on extremely vast amounts of data \cite{vaswani-attenion-is}. Numerous studies have empirically evaluated LLMs on complex linguistic tasks including pragmatic and discourse analysis, theory of mind reasoning, and sentiment analysis \cite{settaluri-pub-a,amirizaniani-do-llms}.

In the realm of HCI research, there has been a growing interest in evaluating how LLMs can be responsibly utilized for assisted writing. Newman et al. explored the trade-offs of providing sentence-level suggestions as opposed to message-level suggestions \cite{newman-comparing-sentence}, and the Wordcraft project embedded LLMs within a text editor to assist storytellers, helping them overcome writer's block and sparking creativity \cite{andy-wordcraft-a}. Additionally, Goodman et al. developed LaMPost, a browser plugin that employs LLMs to aid dyslexic users in crafting and revising emails, including generating email headers and previewing email content \cite{goodman-lampost-design}, and Jang et al. explored the use of LLMs by autistic adults in the workplace as an alternative to seeking social communication support from coworkers, friends, and family \cite{jiwoong-its-the}. While our work expands on this line of research, HCI researchers have also focused on creating guidelines for responsible AI usage, and reducing the risks, biases, and ethical concerns with providing AI-assistance to humans \cite{lima-human-perceptions, yildirim-investigating-how, weisz-toward-general, blodgett-responsible-language}.

\subsection{Autism and Computing}
Human-Computer Interaction researchers have been instrumental in developing new social support tools \cite{tartaro-virtual-peer, washington-a-wearable} and therapeutic interventions \cite{gesture-therapy, jeong-lexical-representation} for autistic individuals. Some of these techniques utilize simulations to create a controlled environment that enables users to safely explore and learn how to handle scenarios that could be challenging or risky in real life. For instance, Park et al. integrated an augmented reality (AR) interface with drama therapy to offer effective, universal, and accessible language therapy to autistic children \cite{park-aedle-designing}, and Boyd et al. employed virtual reality (VR) to support proximity regulation for autistic individuals \cite{boyd-vrsocial}. Prior research has also shown that use of technology promotes higher engagement among users \cite{tarantino-on-the}, can be less resource-intensive compared to conventional therapy \cite{washington-superpowerglass}, and allows for customization that better addresses individual needs \cite{dawson-behavioral-interventions}, moving away from a one-size-fits-all approach. 

Given that autism varies greatly from person to person, the level of customization existing tools provide falls short of what is truly needed, underscoring the need for rapidly adaptive systems capable of supporting real-time content generation \cite{ahsen-designing-a}. In our study, we explore if recent advances in generative AI \cite{jacob-bert-pretraining, gpt4-openai-and} can meet these customization needs in the specific context of AI-assisted instant messaging. In addition, most tools aim to help autistic users in enhancing their social skills, overlooking the collaborative nature of social interactions. We discuss how social interactions can be conceptualized as fundamentally collaborative, and how this observation motivates distributing the responsibility to reduce communication breakdowns across all users.



\begin{figure*}[t]  
    \centering
    \pdftooltip{\includegraphics[width=\textwidth]{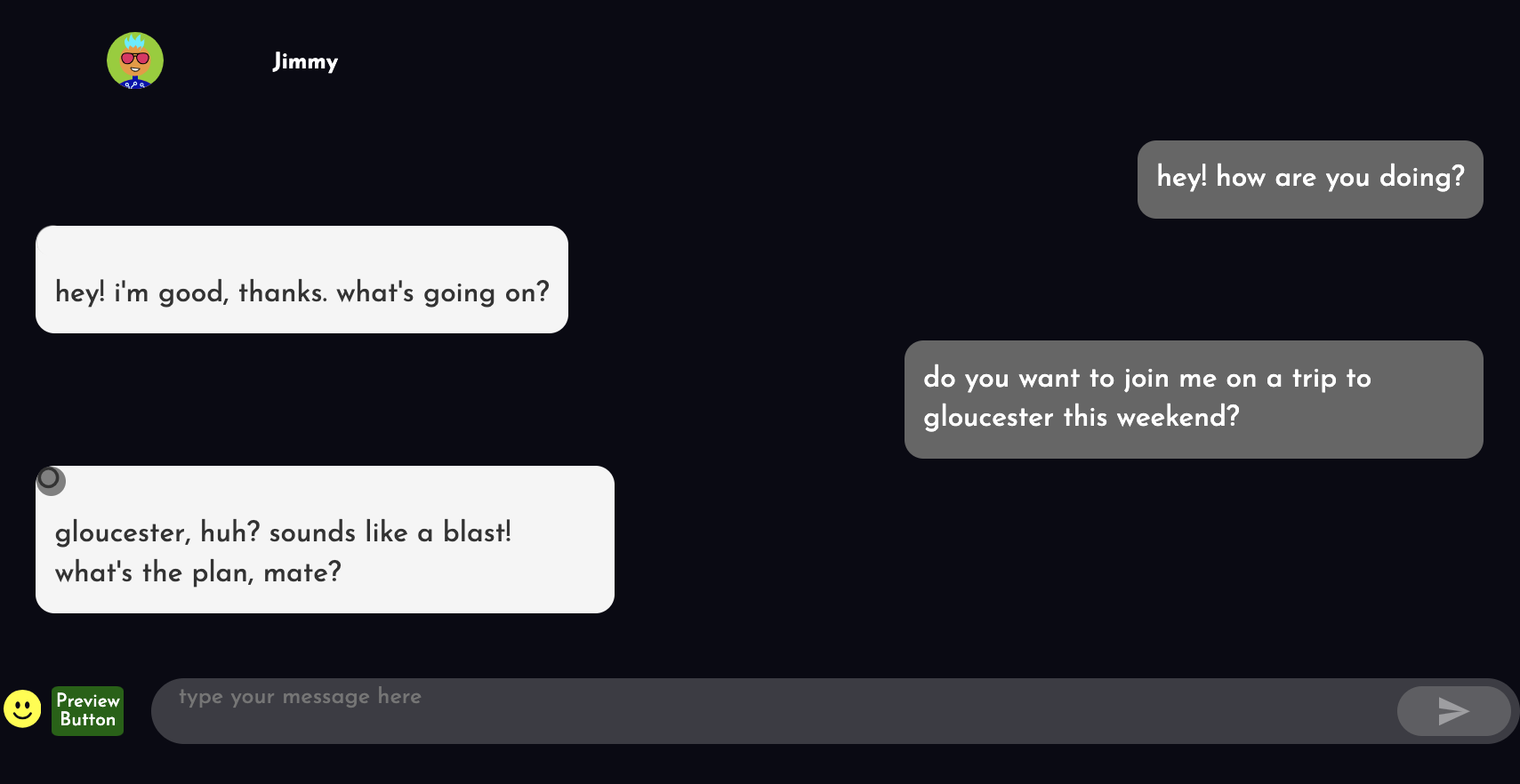}}{}
    \Description{This figure shows a screenshot of the layout of TwIPS, with a conversation in progress. At the top left, there is a picture featuring an avatar with the name "Jimmy" next to it. The conversation in this screenshot begins with a message from a hypothetical participant, saying, "hey! how are you doing?", to which Jimmy replies on the left with, "hey! i'm good, thanks. what's going on?" The next message from the participant on the right is, "do you want to join me on a trip to gloucester this weekend?". Jimmy responds on the left, "gloucester, huh? sounds like a blast! what's the plan, mate?" 
    
    At the bottom of the interface, there is a text input field with the placeholder text "type your message here" and three distinct buttons. On the left of the text input field is the preview button and an emoji to toggle the emoji menu, and on the right, there is a send button represented by an arrow icon. The background of the interface is dark, and the chat bubbles are contrasting, with Jimmy's in white and the participants in a darker shade. In this way, the layout of TwIPS mirrored a standard, modern instant messaging application.}
    \caption{UI of the TwIPS prototype.}
    \label{fig:layout}
\end{figure*}

\section{Overview of T\MakeLowercase{w}IPS}
\label{sec:prototype-overview}
This section motivates the design of TwIPS and describes each of its three features.

\begin{figure*}[htp] 
    \centering
    \begin{subfigure}[t]{0.48\textwidth}
        \centering
        \pdftooltip{\includegraphics[width=\textwidth]{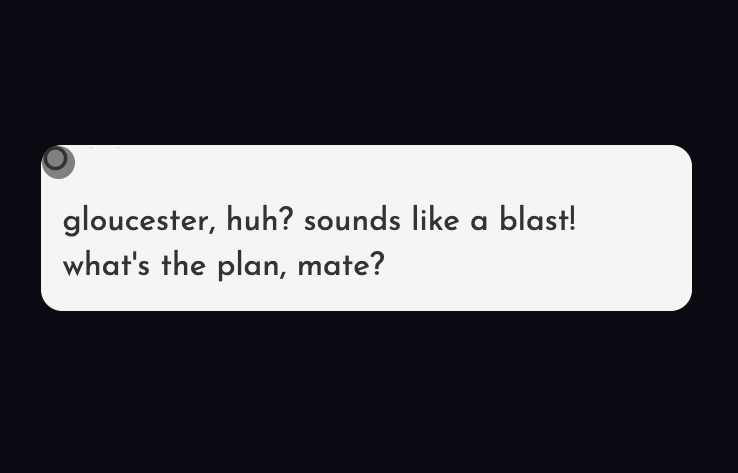}}{}
        \Description{This figure is a screenshot of a close-up of a single message within TwIPS' layout. The chat bubble is white with black text inside it and is positioned against a dark background. The text reads, "gloucester, huh? sounds like a blast! what's the plan, mate?" There's a small, circular gray symbol in the upper left corner of the message bubble, indicating that this message contains potentially ambiguous language elements. The symbol serves as a marker for received messages that might need the receiver's further attention. This enabled us to subtly highlight messages that could be potentially ambiguous.}
        \caption{Messages containing at least one ambiguous language element are marked with a gray symbol in the upper left corner of the chat bubble.}
        \label{fig:sub1}
    \end{subfigure}
    \hfill 
    \begin{subfigure}[t]{0.48\textwidth}
        \centering
        \pdftooltip{\includegraphics[width=\textwidth]{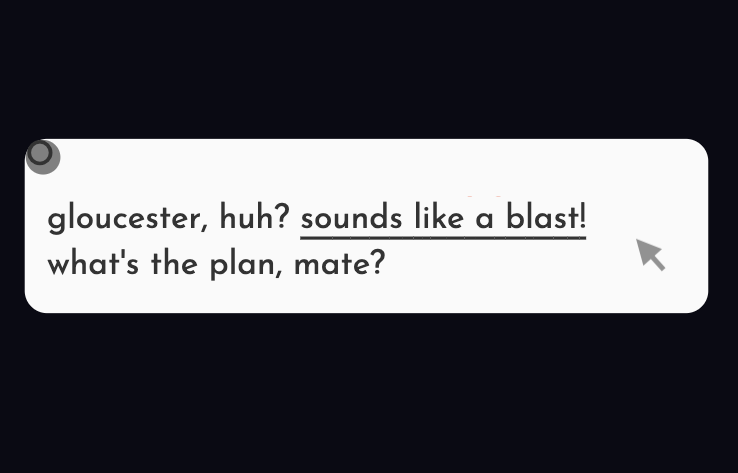}}{}
        \Description{
        This image features a close-up of the same text message as in the previous figure. The message bubble is white with black text, and it's set against a dark background. The text inside the message bubble reads: "gloucester, huh? sounds like a blast! what's the plan, mate?" A section of the text, "sounds like a blast!" is underlined, indicating that this phrase has been identified as potentially ambiguous. A mouse cursor appears to the right of the underlined text within the message bubble, indicating that the underlining is a result of hovering the cursor over the message bubble. In this way, hovering the cursor over any message bubble containing the gray circular symbol (indication potential ambiguity) underlines the ambiguous element(s).}
        \caption{Hovering the cursor over the chat bubble underlines the ambiguous element(s).}
        \label{fig:sub2}
    \end{subfigure}
    
    \vspace{2em} 

    \begin{subfigure}[b]{0.48\textwidth}
        \centering
        \pdftooltip{\includegraphics[width=\textwidth]{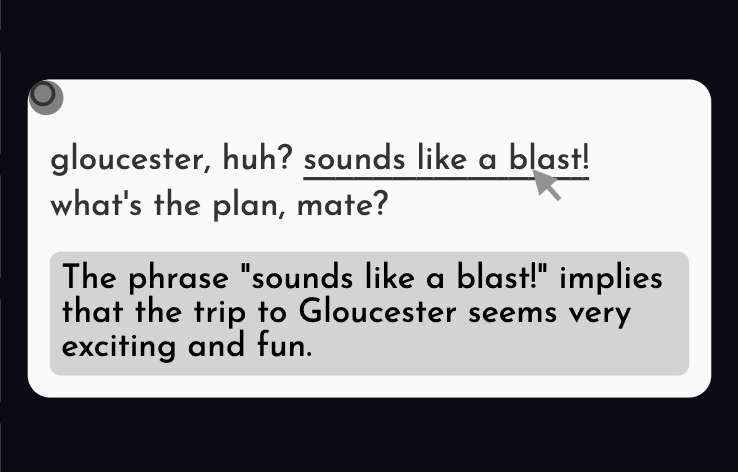}}{}
        \Description{This figure features a close-up of the same text message as in the previous two figures. The message bubble is white with black text, and it's set against a dark background. The message reads, "gloucester, huh? sounds like a blast! what's the plan, mate?" The phrase "sounds like a blast!" is underlined, and a cursor hovers over the underlined text, showing that the user has clicked on the underline phrase. Below the underlined phrase is a smaller, secondary text bubble within the main text bubble containing an explanatory note which reads: "The phrase 'sounds like a blast!' implies that the trip to Gloucester seems very exciting and fun." The explanatory note's bubble has a lighter background and rounded corners. In this way, users could click on any language element that was underlined and receive an explanation for it.}
        \caption{Clicking on an underlined segment triggers an explanation to appear.}
        \label{fig:sub3}
    \end{subfigure}
    \hfill
    \begin{subfigure}[b]{0.48\textwidth}
        \centering
        \pdftooltip{\includegraphics[width=\textwidth]{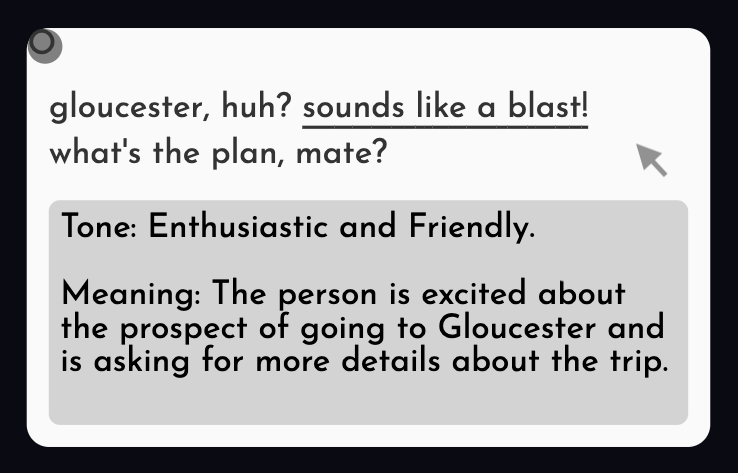}}{}
        \Description{The second figure features a close-up of the same text message as in the previous three figures. However, this time, the user has clicked on a non-underlined segment of the message bubble. As a result, below the main message is a secondary text bubble, which provides a brief description of the overall tone and meaning of the message. It reads, "Tone: Enthusiastic and Friendly. Meaning: The person is excited about the prospect of going to Gloucester and is asking for more details about the trip." In this way, by clicking anywhere in the message bubble other than the underlined language elements, users could get an overall tone and meaning of any message.}
        \caption{Clicking on anywhere else in the chat bubble displays a description of the message’s overall tone and meaning.}
        \label{fig:sub4}
    \end{subfigure}

    \caption{{\sc{Interpret}} in Action.}
    \label{fig:interpret}
\end{figure*}

\begin{figure*}[t] 
    \centering
    \begin{subfigure}[b]{\textwidth} 
        \centering
        \includegraphics[width=\textwidth]{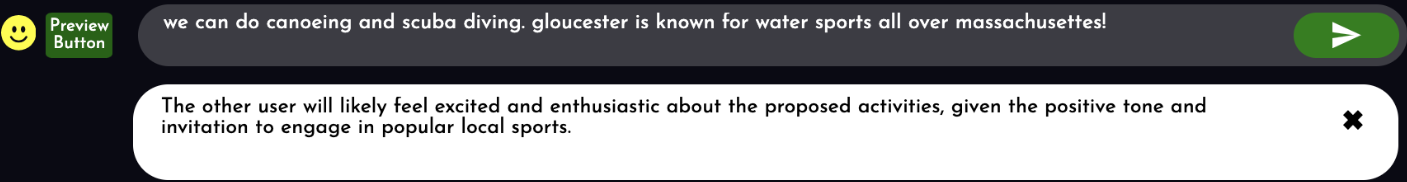}
        \Description{This image is a close-up screenshot of the message writing box and its surrounding user interface (UI) elements, showing the Preview feature in action. On the left, there's an emoji button for users to toggle the emojis menu, and another button labeled "Preview Button". On the right,there is a green send button with an arrow icon. The central part of the image shows a text input area where a user has typed this message to send: "we can do canoeing and scuba diving. gloucester is known for water sports all over massachusetts!" Below this message, a secondary, lighter text box contains a feedback generated by the Preview feature that reads, "The other user will likely feel excited and enthusiastic about the proposed activities, given the positive tone and invitation to engage in popular local sports." At the far right, there's a cancel or close button with an "x" icon, to hide the text box containing the feedback. In this example, the message and
        preview's feedback were fairly positive, hence Suggest was not triggered and an alternative message was not produced.}
        \caption{\textsc{Preview} describes how the recipient of the message may feel upon reading the user’s message. In this example, the message and preview were fairly positive, hence \textsc{Suggest} was not triggered and an alternative message was not produced.}
        \label{fig:sub21}
    \end{subfigure}

    \vspace{2em} 

    \begin{subfigure}[b]{\textwidth} 
        \centering
        \includegraphics[width=\textwidth]{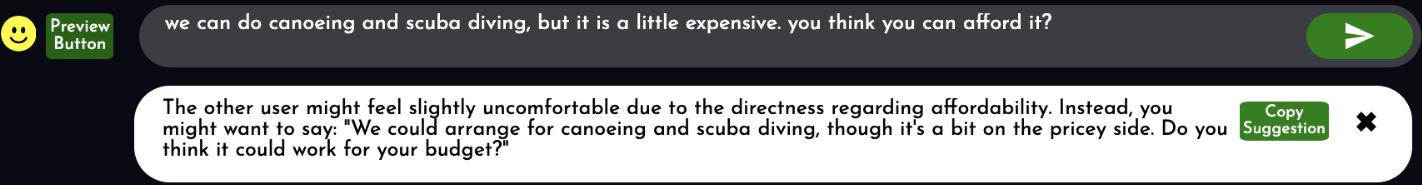}
        \Description{
        This image is a close-up screenshot of the message writing box and its surrounding user interface (UI) elements, showing the Preview and Suggest feature, both, in action. On the left of the message writing box, there's an emoji button for users to toggle the emojis menu, and another button next to it labeled "Preview Button". Users can click on this button to toggle the Preview feature. On the right, there is a green send button with an arrow icon. The central part of the image shows the message writing box where a user has typed the following message to send: "we can do canoeing and scuba diving, but it is a little expensive. you think you can afford it?" Below this message writing box, a secondary, lighter text box contains the feedback generated by the Preview feature corresponding to that message, which reads: "The other user might feel slightly uncomfortable due to the directness regarding affordability. Instead, you might want to say: "We could arrange for canoeing and scuba diving, though it's a bit on the pricey side. Do you think it could work for your budget?"". Within the secondary text box to the right, there is a green button labeled "Copy Suggestion" for easy use of the alternative message suggested, and a cancel or close button next to it with an "x" icon, that users can click to hide the secondary text box.}
        \caption{In this example, a negative preview and an alternative message were produced by \textsc{Preview} and \textsc{Suggest}, respectively.}
        \label{fig:sub22}
    \end{subfigure}

    \caption{{\sc{Preview}} and {\sc{Suggest}} in Action.}
    \label{fig:preview-suggest}
\end{figure*}

\subsection{Motivation}
The design of \textsc{Interpret}, \textsc{Preview} and \textsc{Suggest} is inspired by challenges and concerns raised by autistic individuals, in prior work, related to expressing and interpreting emotions and intent online \cite{barros-my-perfect, page-perceiving-affordances}. These challenges include frequently misinterpreting others, struggling to understand nuanced language, and difficulties in making oneself understood, which forces users to repeatedly read and revise what they want to write and align it with neurotypical communication styles. Without intensive proofreading, they could come off as blunt to others and trigger harsh reactions. These studies also highlight concerns over the extensive use of emojis and GIFs in text-based communication, as both carry nuanced, context-dependent meanings, and emojis are stylized versions of facial expressions which are hard to comprehend for many autistic individuals. Participants in our study reaffirmed many of these challenges.
\subsection{Design and Functionality}

The user interface (UI) of the TwIPS prototype resembles the layout of standard chatting applications, featuring separate chat bubbles for received and sent messages, a text input box, the recipient's name, and buttons for sending messages and accessing emojis, as shown in Figure \ref{fig:layout}. Participant's messages are displayed on the screen's right side in dark-grey chat bubbles, and incoming messages are shown on the left side in light-grey chat bubbles. Additional UI elements to support \textsc{Interpret}, \textsc{Preview} and \textsc{Suggest} are described in detail in the subsequent subsections below. \textsc{Interpret} is designed to help users in understanding messages from others, whereas \textsc{Preview} and \textsc{Suggest} are intended to assist them in composing their own messages.

\subsubsection{\textsc{Interpret}} This feature, as implied by its name, is designed to help users comprehend others' messages. It serves two functions: a) it describes the overall tone and meaning, and b) it individually identifies and explains ambiguous language elements, such as sarcasm, metaphors, and emojis, in incoming messages. A message containing ambiguous language elements is marked with a grey circular symbol in the upper left corner of its chat bubble, and users can hover over any chat bubble with this symbol to reveal the underlined ambiguous language element(s), as shown in Figures \ref{fig:sub1} and \ref{fig:sub2}, respectively. Clicking on an underlined language element expands the chat bubble to reveal its meaning, and clicking anywhere else on the chat bubble expands it to show the overall tone and meaning of the message, as shown in Figures \ref{fig:sub3} and \ref{fig:sub4}, respectively. Only one type of explanation, either for ambiguous elements or the overall message, is displayed at a time.

\subsubsection{\textsc{Preview}} 
This feature is intended to help participants compose messages. It serves two primary functions. First, it enables users to preview recipients' likely emotional reaction to their message, helping them ensure the emotional tone of their message comes across as intended to the recipient and make adjustments to it if needed. Users can toggle it through the `Preview' button positioned next to the emoji button. After toggling \textsc{Preview}, a new text box appears below the message writing box to display the preview, which can be positive or negative, as shown in Figures \ref{fig:sub21} and \ref{fig:sub22}, respectively. The new text box disappears automatically once the user clicks on the `Send' button or the cross symbol present on its far-right side. Second, \textsc{Preview} flags messages as \textit{blunt} or \textit{not blunt}, whenever the user clicks the `Preview' or `Send' buttons. In the former scenario i.e., a message is flagged as blunt, it generates a negative preview and triggers \textsc{Suggest} to generate an alternative message. If the user still wants to send their original message, they can click on the `Send' button again to bypasse \textsc{Preview}'s flagging functionality. 
\subsubsection{Suggest} This feature complements \textsc{Preview} by generating a differently phrased but softer, or \textit{less blunt}, alternate message for any message flagged as blunt by \textsc{Preview}, while preserving the intent of the user's message in the alternative. The suggested message is appended to the feedback provided by \textsc{Preview}, as shown in Figure \ref{fig:sub22}. Users can click on the `Copy Button' to copy the suggestion to the message writing text box. \textsc{Preview} and \textsc{Suggest} work in conjunction with one another, with \textsc{Preview} providing an explanation as to why the user should consider the suggested alternative. \textsc{Preview} bases its explanations off the most likely reaction/perspective of the recipient, with the hope that this will make them sound less corrective and more subjective as opposed to sounding instructional and rigid, ultimately leaving the decision to take or leave the suggested alternative up to the user.

\subsection{Prompting Strategy}
Writing effective prompts requires thorough testing and iterative refinement to elicit dependable and accurate responses from the LLM. During development, we experimented with various prompts before finalizing the ones employed in our final prototype. The prompt templates used for each of the three features in TwIPS are provided in Appendix \ref{sec:appendix-a}. We used GPT4's code API (version: 1106-preview) through Microsoft Azure \cite{microsoft-whats-new}. We applied few-shot learning to our prompts where possible, which is a well-known technique to design custom prompts by incorporating several examples of a task to guide a language model as it performs that task \cite{brown-language-models, openai-prompt-engineering, goodman-lampost-design}. Additionally, we appended the entire conversational history to all prompts, ensuring the LLM had a comprehensive understanding of the conversation's context. We also made feature-specific modifications to prompts where needed. In the prompt used with \textsc{Interpret}, we specified to check for emojis, figurative language, and phrases with an indirect meaning. The tone, meaning, and ambiguous language elements for each message were pre-fetched as soon as it was sent, but explanations for ambiguous language elements were fetched only when a user clicked on one. In the prompt used with \textsc{Suggest}, we made it explicit that the writing style and stance taken in the suggested alternative should match the writing style and stance of the user's original message. While we do not claim that these prompts are ideal, they were adequate for the purposes of our in-lab study within a controlled environment as they provided consistent, reliable results.
\balance

\section{Methodology}


\subsection{Recruitment}
Participants were recruited from a university setting in USA, over a period of two months, through flyers posted in various buildings around campus. Interested individuals completed a screening survey to ascertain their eligibility. The inclusion criteria were: a) being aged eighteen years or older b) fluency in English reading and writing c) ability to perform basic computer tasks, and d) having a formal autism diagnosis or self-identifying as autistic. Recognizing disparities in access to diagnostic methods and procedures, we did not require a formal autism diagnosis to be eligible to participate in the study \cite{aylward-racial-ethnic} similar to other recent studies involving neurodivergent populations \cite{goodman-lampost-design, dotch-understanding-noise}.

A total of fourteen participants were recruited, out of which eight enrolled and completed the study. Table \ref{tab:participant_demo} shows information about the participants. Each user-study session lasted approximately two hours. Participants received an Amazon gift voucher worth twenty-five USD upon the study's completion. The study environment was a quiet room with adjustable lighting to accommodate participant needs. Additional accommodations, such as travel assistance or specific lighting requirements, if any, were identified through the screening survey and later addressed. This study adhered to ethical guidelines and was approved by the institition's Social, Behavioral, and Educational Research Institutional Review Board (IRB).

\subsection{User Study Overview}
This sub-section outlines the design of our two-phased user study. The user-study was carefully designed to provide users with conversational scenarios where they could utilize each of the three features in TwIPS, while ensuring user-autonomy, consistency and dynamicity across their experiences. Before each phase, participants received a handout detailing the prototype's features and tasks associated with that phase, along with a demonstration by a member of the research team. Participants were free to ask questions at any time.

\subsubsection{Phase One: A Scripted Conversational Scenario}

\begin{figure*}[t]
    \centering
    \includegraphics[width=0.7\textwidth]{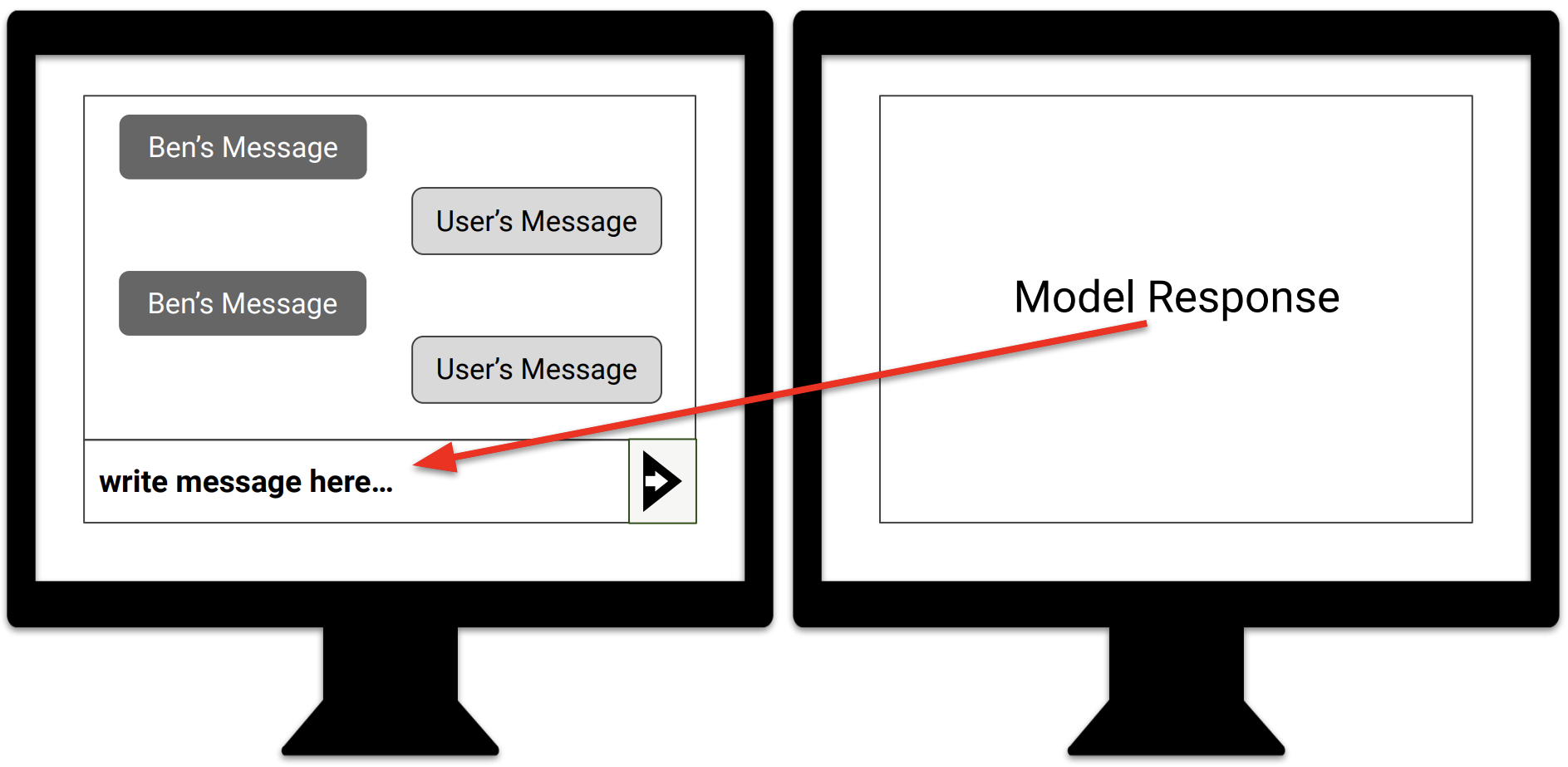}
    \Description{This figures shows the setup for phase 1 of the user-study. 2 computer monitors side by side are shown in the figure; the left screen displays the TwIPS' interface with a conversation showing alternating messages from 'Ben' and the study 'User', and a text input area labeled "write message here…" with a send button next to it. The right screen displays the text "Model Response," indicating where the model-generated response appeared. An arrow points to the text input area on the left screen from the model response on the right screen, showing that participants could see and copy the model response from the right screen to the message writing box. The model response on the right screen was updated each time Ben sent a new message.}
    \caption{In phase 1, participants were provided with two monitors. The TwIPS prototype, used for exchanging messages with Ben, was shown on the left monitor while the model response was displayed on the right monitor. The model response was updated automatically each time Ben sent a new message.}
    \label{fig:user-study}
\end{figure*}

In phase one, each participant engaged in a simulated one-on-one conversation with an imaginary character, Ben. To facilitate the simulation, the research team came up with a scripted conversation between Ben and participants prior to the study. The script centered around planning a birthday surprise for Jack, who was assumed to be a mutual friend of Ben and the participants. The same script was used for all participants and it is provided in Appendix \ref{sec:appendix-b}. Participants' messages in the script were a mix of blunt and non-blunt messages - this was not revealed to them. These served as \textit{model responses} that participants could send to Ben, and were revealed to them one-by-one, each paired with one of Ben's messages, as shown in Figure \ref{fig:user-study}. Participants could either use the model response as-is, modify it, write a new response or choose the alternative suggested by \textsc{Suggest} when available, if they found the model response to be too blunt to send. Participants were given ample time to decide how they wanted to respond to Ben, explain the rationale behind their decision, and thoroughly describe their impressions of the model responses as well as the feedback and suggestions provided by {\sc{Preview}} and {\sc{Suggest}}, respectively. Participants were encouraged to think aloud during phase one.
 
\begin{table*}[t] 
\centering
\begin{tabular}{llllll}
\hline
\textbf{P\#} & \textbf{Diagnosis} & \textbf{Self-Identifying} & \textbf{Age} & \textbf{Gender} & \textbf{Texting Usage} \\ \hline
\textbf{P1}  & Y & Y & 25-34 & Female & Very Frequent \\
\textbf{P2}  & Y & Y & 18-24 & Female                  & Very Frequent \\
\textbf{P3}  & Y & Y & 18-24 & Non-binary/third gender & Very Frequent \\
\textbf{P4}  & N & Y & 35-44 & Female                  & Very Frequent \\
\textbf{P5}  & N & Y & 18-24 & Male                    & Very Frequent \\
\textbf{P6}  & N & Y & 18-24 & Non-binary/third gender & Very Frequent \\
\textbf{P7}  & Y & Y & 18-24 & Male                    & Very Frequent \\
\textbf{P8}  & N & Y & 18-24 & Female                  & Very Frequent \\ \hline
\end{tabular}
\Description{
This table shows information about participants. P1, P2, P3, and P8 had received a formal autism diagnosis while the rest 4 participants self-identified as autistic. P1 was aged 25-34, P4 was aged 35-44 and all other participants were aged 18-24. P1, P2, P4 and P8 identified as female. P3 and P6 identified as non-binary/third gender, and P5 and P6 identified as male. All participants reported their texting usage was "very frequent".}
\caption{Information about participants.}
\label{tab:participant_demo}
\end{table*}
We imposed a number of constraints as part of phase one's design. First, we concealed the `Preview' button from participants, so {\sc Preview} was set to activate only automatically in phase 1 i.e., if a participant was blunt in their response to Ben, {\sc Preview} might flag and prevent it from getting sent. Second, we intentionally included several blunt messages in the set of model responses provided to participants. Third, participants were instructed that if they wanted to modify the model response, or write a new response, they should do so in a manner that might alter its wording/sentence structure but maintain its semantics/stance. This design let us examine, in an in-lab setting, how our participants prefer to communicate in situations demanding directness or bluntness, and their reactions to being nudged by \textsc{Preview} and \textsc{Suggest} to revise their messages. At the same time, it granted participants the freedom to modify or rewrite their messages, decide whether to take up a suggestion or ignore it, and thoroughly explain the rationale behind their decisions.

\subsubsection{Phase Two: A Nuanced AI-based Simulation}
In phase two, participants initiated a second conversation with Ben with the goal to plan a trip with him to Gloucester, a waterfront city in Massachusetts. To aid participants in initiating the conversation, background information about notable destinations in Gloucester was provided to them. Participants were encouraged to compose messages in their own, unique writing styles. To simulate a more realistic and dynamic conversation, Ben was configured to be an AI-persona whose responses were dynamically generated on the fly via GPT4 in phase two. 

Additionally, GPT4 was configured through prompts to introduce ambiguous language elements such as positive sarcasm, figurative language, emojis and jokes in its responses. This ensured that Ben's AI-persona acted in a way that prompted participants to use \textsc{Interpret}, thereby testing its capability to identify and explain ambiguous language elements. Contrasting with phase 1, where {\sc Preview} was set to activate only automatically, phase 2 offered participants the option to manually toggle it through the `Preview' button, which they could click before sending a message. This enabled us to observe how users interact with {\sc Preview} and {\sc Suggest} when given more control. Participants were encouraged to explore and experiment with all 3 features to gain a holistic understanding of the application's features. In both phases, participants were informed that Ben was not a real human to allow them to stress our prototype with all kinds of messages without the fear of actually hurting him.



\subsection{Data Collection and Analysis}
Upon completing both phases of the study, participants took part in a semi-structured interview followed by a survey in the same sitting. The post-study semi-structured interview delved into participants' perceptions of TwIPS' usefulness and suggestions for improvement (based on how it helped or did not help them during the study) and their experience with traditional texting applications. Following the interview, participants completed a survey comprising 19 Likert-scale questions. Each question was rated on a 7-point scale, ranging from 'Completely Disagree (1)' to 'Completely Agree (7)', with an additional option available for questions where standard options were not applicable. 
We adapted our survey design from the survey used by Goodman et al \cite{goodman-lampost-design}. The survey results provided a holistic overview of user perceptions concerning each feature as well as the prototype as a whole, and qualitative insights helped explain them in detail. The primary themes in the survey included the prototype's usefulness, correctness, and impact on participants' sense of self-autonomy. Throughout the duration of the study, participant's audio and screen were recorded for later analysis. Written consent was obtained from each participant before the start of the study.

Given our small sample size (N=8), we state descriptive statistics with survey results, such as average and standard deviation, along with the verbatim survey question. To conduct qualitative analysis, we used Braun and Clarke’s approach to thematic coding \cite{braun-using-thematic} using a deductive approach. Prior to the study, we developed a set of deductive codes to categorize: existing challenges with texting; communication style and preferences for self-expression and interpretation in text; positives, negatives, and improvements for each feature; and perceptions of self-autonomy and accessibility. A member of the research team transcribed the data and contextualised the transcripts with additional information from screen recordings. After importing the transcripts in NVivo, they extracted relevant quotes for each code, grouped these into themes, discussed the themes with other team members, and then reviewed and refined them. A second member, who was not part of the initial study team, validated the themes and the data associated with each theme.
A similar approach was employed by Ahsen et. al \cite{ahsen-designing-a}.

\section{Findings}

In this section, we present background on participants' communication styles and preferences, their reactions to the TwIPS prototype, and their suggestions for improving it. A summary of participants' qualitative feedback on the TwIPS prototype is provided in Table \ref{tab:feature_summary}.

\subsection{Communication Styles, Practices and Preferences}
\subsubsection{Sustaining One's Weight}
Most participants, including P1, P2, P3, P4, P5, P6, and P8, demonstrated a strong inclination to maintain their own weight in the conversation. In addition to making affirmations or negations to Ben's questions (Ben was the AI character involved in the user study), participants made a conscious effort to pose their own questions and steer the conversation. In instances of disagreement or while expressing disinterest towards Ben's suggestions, they proposed alternative suggestions to prevent the burden of generating new ideas from falling on Ben. At one such instance, Ben asked P4 if they knew Jack's (Jack was assumed to be a mutual friend of Ben and the participant) birthday was coming up, and P4 remarked, \textit{ "A response like `Yeah, I know!' will shut down the conversation and put the burden on Ben to decide what to talk about next. I'll add `Should we do something [to celebrate Jack's birthday]?' to my response"}. P8 resonated with this remark. Moreover, P1 highlighted the struggle to maintain balance in text-based conversations is a common challenge faced by autistic individuals, \textit{"Maintaining a balance in text messages is difficult. We often respond directly to the question posed without contributing equally to the conversation's flow. When I talk to people who are not autistic, their biggest complaint when texting autistic people all the time is: I feel like they do not want to talk to me because they only answer my question but never continue the conversation"}. In this way, a key aspect of participants' thought process involved recognizing that simply disagreeing without offering alternatives, or responding to questions without prompting further discussion, might shift the burden of maintaining the conversation's momentum onto others.








\subsubsection{Clarity and Directness}
Numerous participants, specifically P1, P3, P4, P5, P6, P7 and P8, shared that they strive to text in a clear and direct manner, with the aim of minimizing the chances of being misinterpreted by others. P4 emphasized it was necessary to communicate clearly and directly, particularly with other neurodivergent individuals, as even slight ambiguities could result in them getting caught in a non-stop cycle of overthinking. For P5, being straightforward and concise reduced the chances of misinterpretation in general too, \textit{"...the fewer words you use and the more to the point it [the message] is, there are less extra things for people to misinterpret."} Conversely, P4, P6, and P2 asserted being direct might not be suitable in certain scenarios, like interactions with unfamiliar people or in the midst of disagreements. When P4 wanted to express disinterest in Ben's suggestion, they explained, \textit{"Since I don't know Ben very well, I'm not going to push hard on his expressed preference. If it was someone I knew well, I might be more blunt."} 
In this way, participants highlighted the importance of clarity and directness in texting, emphasizing that the level of directness must be adjusted according to the situational context for effective communication. 



\subsubsection{Masking in Text}

Masking entails consciously or unconsciously altering one's behavior to conform to societal expectations -  autistic individuals often feel the need to mask while interacting with neurotypical individuals. Our participants explained the different factors they consider and strategies they use to effectively mask in text. P3 believed it was necessary to mask with strangers until a reasonable level of familiarity was reached, at which point they could reveal their true communication style without fear of judgement. For P4, masking needed to be done more pervasively to adhere to social norms, and P1 rewrote any message they needed to mask. This was a flexibility afforded by texting, unlike in FTF interactions where words are spoken without the opportunity for revision, \textit{"I typically start typing out my response, and then if I realize that I need to mask, I end up fully rewriting it. It’s the same idea of like turning everything into sounding nice even when you aren’t being nice"}. P4 likened masking to a ``transparency slider'', explaining how they adjust it to reveal varying degrees of their personality in different contexts or selectively display certain aspects of it. P4 and P7 stressed on the significance of maintaining authenticity while masking, with P4 stating,
\textit{"If you just copy somebody with your masking, then you look like somebody else, and people will always interact with you that way. Whereas, if you make it yours, but more like a translator than a cover, then it's more like you, except understandable to people who are not"}. This illustrates that autistic users mask to conform to the norms of text-based conversation, which can differ significantly from those of FTF conversations, and make a concerted effort to preserve the authenticity of their writing style.

\subsubsection{Punctuation and Crutches}

P1, P2, P3, P4, P5 and P7 pointed out that punctuation plays a key role in tone interpretation and the personalization of one's writing style. P1 associated imperfect punctuation and spelling with an amicable or colloquial tone, and linked precise punctuation, such as using a single question mark instead of several, to seriousness and stronger emotions like bluntness or annoyance. For P2 and P5, exclamation points typically signalled positivity and excitement, unless the message inherently carried a negative meaning. P4 and P7 consistently capitalized the first letter of each word in the message and avoided slang, as they believed that using non-standard words and incorrect punctuation made the messages sound unlike their authentic self. P4 expressed, \textit{"That doesn't sound like me. So, I would never, ever send a message that had non-existent words"}. Furthermore, P1 and P2 employed conversational aids to more accurately express their tone. P2 mentioned that sometimes they include a disclaimer when feeling too fatigued to invest the effort required to articulately phrase their message. The disclaimer was basically a follow-up message, requesting the recipient to interpret the previous message in the specific tone intended by the sender. Similarly, P1 noted they frequently include extra `context' to explain their punctuation choices. By `context', P1 meant adding tone indicators like `/sarcastic' and `/excited' at the end of their messages to clearly convey the intended tone, \textit{"I often put a lot of context in my message, because messaging people is very stressful... if I use a specific punctuation mark, I'll explain its purpose"}.

\subsubsection{Resolution Strategies}
Based on past experience, participants reported it was challenging for them to discern tone and meaning within their own text messages, as well others' messages. To counteract this, they adopted a range of strategies. P1 shared that sometimes, they seek help from their close friends before sending out a message or email, \textit{"I will send it [a screenshot] to people and be like: does this sound okay? And they'll be like: that sounds like you're heavily critiquing that person. You should definitely change xyz things. So, it takes me a long time, because I have to rewrite a lot of what I say."}. For P3 and P6, vocalizing other's messages enhanced their understanding of the intended tone. P2, P3 and P7 preferred to inquire the sender directly to avoid making any assumptions or misinterpretations about their tone, while P5 felt it was better to resolve ambiguities in person, expressing worry that trying to clarify something over text could further add to the confusion, \textit{"I'm uncertain if Ben is being passive-aggressive. Confronting him might just complicate things if he's not actually being that way. So, I won't bring it up, at least not over text"}. 






\begin{table*}[h!]
\centering
\begin{tabular}{|p{1.4cm}|p{3.7cm}|p{5.5cm}|p{5cm}|}

\hline
\textbf{Feature} & \textbf{Description} & \textbf{Participant Reactions} & \textbf{Suggested Improvements} \\ \hline
\textsc{Interpret} & Describes the overall tone and meaning of a received message, and identifies and explains any ambiguous language elements in it individually. & 
More nuanced than tone indicators, clarifies use of context-dependent emojis and phrases (but too many underlined parts can get overwhelming), and useful when asking for clarification by the sender isn't possible. Benefits neurotypical users as no need for them to constantly explain themselves now. & 
A visual measure of \textsc{Interpret}'s confidence in its responses could help users establish an appropriate level of trust in it. The interpretation of the sender's message shown to the receiver should be visible to the sender, allowing for corrections if needed.

\\ \hline
Preview & Describes the recipient's likely reaction to a message before it is sent, ensuring it comes across as intended by the sender. & 
Facilitates reflection on writing technique and style, eliminates the need to ask feedback from others, and provides guided assistance to improve a message. Automatic activation leads to a sense of scrutiny but reduces the tendency to overthink.
& 
Option to preview tone/meaning of a message before it's sent, expanding the flagging criteria, tailoring feedback for different age levels, and proactively nudging users to contribute to a conversation could increase its usefulness.
 \\ \hline
Suggest & Suggests alternate messages to the sender, while ensuring the intent of their original message remains preserved but the alternative has a softer tone. &  
Suggestions were softer, more thoughtful, and less imposing, though sometimes the emotional intensity was reduced more than necessary. Did not always match the sender's writing style or fit the conversation's flow.
& 
Instead of a complete sentence, it could suggest an incomplete one and leave the personalizable parts of it for the user to complete. A user-specific calibration process in the start could help enhance personalization.
\\ \hline
\end{tabular}
\Description{This table summarizes the qualitative feedback and suggestions for improvements provided by participants for each of 3 features in TwIPS.}
\caption{Summary of Participants' Qualitative Feedback on TwIPS' features.}
\label{tab:feature_summary}
\end{table*}

\subsection{Feedback on the TwIPS Prototype}

\subsubsection{Reactions to \textsc{Interpret}} 

\paragraph{Seeking Clarification}

For multiple participants, interpreting tone and meaning in text was challenging and often necessitated clarification from the sender. In instances where asking for clarification was not possible, participants expressed \textsc{Interpet} could be incredibly helpful. Specifically, P3 underscored its utility in group chats and on dating apps. \textit{"In group chats, when you're not part of the ongoing conversation, it's hard to tell if you're missing context"}, and \textsc{Interpret} could help with clarifications without directly inquiring the sender in front of other groups members. As a user of dating apps, P3 identified flirting over text as a challenge, \textit{"flirting depends on subtlety and one-on-one responses - the difficulty lies in knowing when to shift the conversation or topic"}. They emphasized \textit{"first impressions are basically everything"} in this context, and \textsc{Interpret} could play a pivotal role in helping users understand innuendos and ensuring responses align with the intended tone, thereby helping them \textit{"say the right thing"} and navigate the \textit{"make-or-break"} nature of initial interactions more effectively. Similarly, P5 and P2 felt that \textsc{Interpret} could be beneficial in the initial stages of talking to or befriending strangers. Perhaps, P1 believed \textsc{Interpret} could potentially benefit both neurotypical and autistic users, explaining \textit{"people who don't feel comfortable asking for clarification could benefit from it, as well as neurotypical individuals, for whom having to constantly explain their tone can get annoying"}. Overall, participants strongly agreed (avg. = 6, s.d = 1.07) with the statement, "The application's interpret feature enabled me to better understand the overall tone and meaning of messages" in the post-study survey.


\paragraph{\textsc{Interpret} or Tone Indicators} Some participants compared \textsc{Interpret} to using tone indicators. Tone indicators are short abbreviations used at the end of a message to clarify the sender's emotional tone, helping the receiver understand the intended sentiment, such as seriousness, sarcasm, or humor. For example, "/s" indicates sarcasm, while "/j" denotes joking. P6 described tone indicators as \textit{"pretty vague because they are limited to a small number of high-level adjectives like `sarcastic' or `joking'"}, and liked how \textsc{Interpret}, on the other hand, is more nuanced as it \textit{"uses multiple, very specific adjectives"} to describe tone. For P6, one's understanding of tone significantly influenced their interpretation of the message's meaning. Since \textsc{Interpret} not only provided tone but also the meaning, it made this connection clearer; for this reason, they believed \textsc{Interpret} surpassed the clarity provided by tone indicators. While P2 might not utilize \textsc{Interpret} for every message, they saw its value in moments where they would want to double-check their understanding, particularly when dealing with sarcasm or metaphors. 

\paragraph{Ambiguous Language Elements}

In addition to clarifying the overall tone and meaning, \textsc{Interpret} was designed to identify and explain ambiguous language elements, such as figurative phrases and emojis. P2 appreciated how the two features were subtly integrated, \textit{"the underlined phrases only appear when I hover over the message, rather than bombarding me with them. This makes the experience less overwhelming and more user-friendly"}. P4 found the underlining helpful as \textit{"sometimes it's hard to even identify the emoji"}, however, they expressed concern that if too many components in a message were ambiguous, it would be \textit{"anxiety inducing to have all of them underlined"}. For P2, P6, and P8 the ability to delve into specific phrases added value in its own right, because one could grasp the gist of the message but might lack understanding of certain phrases within it. P2 argued \textit{"the more information provided, the better, as there might be instances where I grasp the overall meaning but not specific phrases or emojis in the message, such as `pops on the beach' [a phrase Ben used that P2 did not know of]"}. 

Moreover, P2, P3, P4, P6, P7 and P8 saw value in \textsc{Interpret}'s ability to explain emojis. For P2, \textit{"using the skull emoji to indicate laughter rather than the usual laughing emoji can be confusing, so having emoji explanations along with the context in which it's commonly used was really helpful"}. P4 appreciated the range of emojis \textsc{Interpret} covered, and how it handled emojis combinations, \textit{"the range of emojis it can explain is excellent. I like how emoji combinations are explained differently, because the meanings of individual emojis can change when combined"}. Having emoji explanations upfront was convenient for P1, and P4 echoed this sentiment, \textit{"it's possible to search for phrases elsewhere, but having immediate explanations especially for emojis, that are challenging to look up, is very convenient"}. Conversely, P1 and P5 deemed explanations of individual language elements redundant. P1 attributed this to the nature of emojis Ben employed in their interaction, which they found to be self-explanatory, \textit{"Ben seemed to use emojis not as word replacements but more as expressions of excitement, silliness, or perhaps just for the novelty"}. Overall, participants strongly agreed (avg. = 6.5, s.d = 0.76) with the statement, "The application's interpret feature correctly interpreted the overall tone and meaning of messages, taking into account the conversation's context."

\subsubsection{Reactions to \textsc{Preview}}

\paragraph{Facilitating Reflection}
Participants were of the view that \textsc{Preview} facilitated constructive reflection on writing technique and style. P4 identified reflection as a vital part of learning to communicate effectively and independently, \textit{"\textsc{Preview} helps me develop a mental model of what others might find rude, dismissive, or offensive, going beyond resolving a single incident. It's so important to know why something is bad in order to be able to pattern match for future"}. P2 and P6 echoed this sentiment, describing that \textsc{Preview} helped them identify aspects of their messages that could be improved but were overlooked by them. For example, P6 stated, \textit{"\textsc{Preview} made me aware of how I might be suggesting my preferences as the only options without considering Ben's choice"}. Similarly, P2 observed, \textit{"I thought my message was good - but then \textsc{Preview} made me realise what was off"}. P4 expressed that \textsc{Preview} not only helped them find areas for improvement, but also indicated how to make those improvements. 
For instance, P4 found value in the advise to gradually lead into the chat while initializing their chat with Ben, \textit{"guidance on prefacing my main points has been incredibly beneficial. It has prompted me to consider the extent of introductory conversation necessary before diving into the main topic"}. In addition, P1 differentiated her experience as an autistic woman from that of autistic men, \textit{"as women, a lot of our language is already expected to be curved, like our emails have to have exclamation points and, you know, we have to try really hard to sound kind. I feel like this [\textsc{Preview}] would benefit a lot of the autistic men that I know. It would enable useful self-reflection that a lot of autistic men, I think, need. A lot of autistic women are diagnosed later in life, and at that point they have learnt to mask really effectively. So many women, including myself, have mitigated the problems that this is trying to solve in a variety of ways. But I don't know a single autistic man, and I know a good amount, that wouldn't benefit from this"}. P1 further expressed that before sending an email or message, they often obtain feedback on it from others, and \textsc{Preview} was a more convenient way for doing the same thing, \textit{"right now, I require the help of a community [close friends] where I have to send them a screenshot and be like does this sound okay to you?"}. Overall, participants expressed strong agreement (avg. = 6.63, s.d = 0.74) with the statement, "The application's preview feature can help prevent misunderstandings as it enables users to preview recipients' likely response to their message and make adjustments to it before sending."

\paragraph{To Toggle or Not}
In phase 1 of the user study, \textsc{Preview} was configured to activate automatically for blunt messages. In phase 2, participants were given the option to toggle \textsc{Preview} via a button before sending a message, in addition to it activating automatically. Participants held mixed opinions on whether they should have the option to manually toggle \textsc{Preview}. P8 compared \textsc{Preview} with their current approach to effective communication, \textit{"There's a little checklist in my head that I want to run through before I say things. Because I run through that checklist anyway, I don't feel like I lost any autonomy [with \textsc{Preview} automatically doing those checks for me]. So, it fits very nicely into my life"}. P6 highlighted that many autistic individuals may already experience heightened levels of overthinking before sending a message, and introducing the option to manually toggle \textsc{Preview} could exacerbate this tendency and users would be \textit{"bound to overthink more"}. In P6's and P8's opinion, automatic \textsc{Preview} could prevent folks from unnecessarily toggling \textsc{Preview}, as it would automatically activate if an issue was detected. However, P6 acknowledged that manually toggling \textsc{Preview} \textit{"reduces a bit of anxiety because you don't have to hit the send button and then wait for an evaluation, alleviating a lot of pressure since all the revision work happens before you press send"}. Similarly, P7 compared the feeling of constant scrutiny and anxiety resulting from automatic \textsc{Preview} to the sword of Damocles in phase 1 of the study \footnote{In the story of Damocles, a servant envies the king's power but realizes the constant danger the king faces. The king offers the servant the opportunity to experience his power for a day but with the danger of a sword hanging over his head by a single strand of hair, symbolizing the constant risk and responsibility of leadership. \label{damocles:example}}. However, participants noted in phase 2 that manually toggling \textsc{Preview} could also result in positive feedback. P7 and P8 felt \textit{"acknowledged"} and \textit{"cared for"} when \textsc{Preview} described some of their messages as \textit{"understanding", "affirmative"} and \textit{"conscientious"}. Additionally, P2, P3 and P8 saw the value of being able to manually toggle \textsc{Preview} in certain circumstances. P2 expressed, \textit{"if I'm texting my dad, I probably wouldn't use it. However, when I am texting someone I'm not particularly close to, or having a difficult conversation with a friend, I'd likely use it to ensure that I'm conveying my message as intended"}.

\paragraph{Misaligned Feedback}
Occasionally, participants indicated  that although \textsc{Preview}'s judgement was accurate, it was exactly what they wanted to convey, and were unconvinced that they should not send their message. For instance, the following feedback was provided by \textsc{Preview} to P2 when they expressed to Ben a clear preference for not inviting other people to the movies because coordinating with large groups was a hassle: "Your message might be perceived as dismissive by Ben, as it is highlighting a dislike for dealing with large groups." P2 swiftly responded, \textit{"but conveying that is exactly the point of my message"}. Similarly, P3 expressed at one instance, \textit{"I am definitely acting unenthusiastic or uninterested [the words unenthusiastic and uninterested were part of the feedback provided by \textsc{Preview}], which is my goal with this response"}. 
P1, P3 and P6 echoed these sentiments. 
Moreover, P2 pointed out that a message that appears blunt or insulting might be perfectly normal in conversations with a friend that they are accustomed to talking like that. They feared \textsc{Preview's} accuracy \textit{"could be hit or miss depending on the specific dynamics of the interaction"} if it is not able to adapt to those dynamics. Overall, participants valued the option to ignore the feedback if they wanted and continue to send their original message, and disagreed with the following statement: "The application's ability to automatically prevent an inappropriately toned message from being sent negatively impacts user autonomy" (avg. = 2.25, s.d. = 1.98).

\subsubsection{Reactions to \textsc{Suggest}}

\paragraph{Self-expression in the Suggested Alternatives}

Alternate messages produced by \textsc{Suggest} were perceived by participants as softer and more thoughtful, and included justifications with strong opinions or decisions to make them appear less imposing. For instance, when \textsc{Suggest} changed the message "coordinating with others is a hassle" to "organising with a lot of people might complicate our plans, don't you think?", P3 commented, \textit{"the word `hassle' is definitely replaceable. The addition of `don't you think' makes it less confrontational, because it is like we are working together. It is a more tactful way to express the same sentiment"}. Similarly, in phase 1, \textsc{Suggest} changed "Just as long as we don’t go to a seafood restaurant" to "Just a heads up, I’m not
really a seafood fan!", and P6 expressed, \textit{"I [in my original message] come across as if I am imposing a rule, implying that seafood should be completely off the table, whereas the suggestion sounds gentler and more like offering a consideration"}. P2, P4, P5 and P7 expressed similar opinions. 

In certain instances, however, especially in scenarios of disagreement, participants expressed \textsc{Suggest} appeared to undermine their opinions by softening their original message so much that it reduced the intensity and emotional depth more than required. For instance, when \textsc{Suggest} changed the message "umm… is that [inviting others to the birthday party] necessary?" to "what do you think? Should we keep it small or invite everyone?", P4 stated, \textit{"It [the alternative provided by \textsc{Suggest}] does not express the fact that I have a preference [for a small group]. If I took all of the suggestions, without editing any of them, I feel I would have a very passive tone and not be able to give my input. The rephrasing introduces flexibility but at the expense of omitting my preference for a smaller group. I think I'm more likely to be unhappy if I don't ever express my own preferences "}.  P1, P3, P7 and P8 echoed these sentiments, highlighting \textsc{Suggest} hindered their ability to clearly express their preferences. Participants valued the option to ignore the suggestion if they wanted and continue to send their original message, disagreeing (avg. = 2.25, s.d. = 1.98) with the statement, "The application's ability to generate and suggest alternative messages negatively impacts user autonomy."

\paragraph{Need for Personalization}
At many instances, participants felt the suggested alternative messages did not have a human-like writing style, were not personalized to match their own, and used words that did not resonate with the dynamics of the rest of the conversation. For example, when \textsc{Suggest} changed the message "i don’t think we need to [invite others]" to "I guess it could be more intimate if it's just us", P1 commented \textit{"I think it's kind of overdoing it because it's not necessarily supposed to be intimate, right? I think the word choice is kind of bad in this one"}. P4 expressed the alternate messages were often \textit{"AI-ish and pretty generic..."} and P3 found one of the suggestions to be \textit{"a little wordy"}. Overall, participants disagreed (avg. = 3, s.d. = 1.93) with the statement, "It felt as if the message suggestions generated by the application's suggest feature had been written by me." As a result, participants resorted to a number of strategies for utilizing the alternate message in some way instead of using it to replace their original message. For instance, P4 adjusted the alternative suggestion to more closely mirror their own style by matching the word construction to their typical usage, but found it beneficial that the suggested message contained the right content, which they could use without having to come up with it themselves. P3, P4, and P6 chose to selectively integrate phrases from the suggested alternate message into their original message, instead of using the suggested alternative in its entirety. P3 noted that just reading the alternate message also helped them understand what aspects of their original message could be improved.

\subsection{Suggestions for Improvement}

\paragraph{Establishing Trust in \textsc{Interpret}}
P4 emphasized the importance of establishing appropriate trust levels in \textsc{Interpret}, cautioning against over-reliance, \textit{"I don't want myself to think that this can do more than it really can... Now I may over trust it and now we have a problem because I'm going to trust it when I shouldn't trust it"}. They suggested incorporating a visual `certainty measure' to inform users about \textsc{Interpret}'s confidence in the interpretations it made. P4 argued this would be particularly valuable in scenarios involving nuanced elements like slang or inside jokes that could confuse the system itself. In their opinoin, this would help users build a balanced level of trust, encouraging them to also rely on their instincts, \textit{"this might be really valuable in building the right amount of trust so that people know to trust their own instincts too"} and consider the possibility of errors in \textsc{Interpret}'s judgment, \textit{ "having a certainty measure will also cue people to think about the fact that the computer could be wrong"}. 
In addition, P1 and P8 expressed the need to be able to see for senders what was being shown to recipients. P1 expressed, \textit{"it would allow me to see, `Oh yeah, that's about right', and provide the opportunity to correct it if not"}. 


\paragraph{Expanding \textsc{Preview}}
While participants liked the concept of \textsc{Preview}, they believed it could be expanded upon. In P2's opinion, \textit{"it's not just about how the recipient takes your message, but more importantly, it's about ensuring that what you're sending accurately reflects what you intend to say"}. They suggested that, alongside recipient reactions, it would be useful to be able to preview a message's tone and meaning, as this would guarantee that the message conveyed precisely what the sender intended. For P1, who struggled with maintaining their weight in conversations, it was a perfect use-case for {\sc{Preview}} to nudge the user to contribute more to the conversation when needed, in addition to nudging them when a message came across as blunt, \textit {"the app could prompt me to think about how I can continue the conversation..."}. 
Moreover, P4 argued that \textsc{Preview} should be capable of identifying messages that are not inherently rude but may become rude if done repeatedly. They stated, \textit{"It's challenging when actions don't follow a strict rule, and only become rude with repetition. For example, \textsc{Preivew} could direct me to suggest another genre to Ben, which would make for an even better message than asking Ben to pick another genre, especially since I've already said no [to the genre he suggested] like three times"}. Additionally, P5 proposed it would be useful to have \textsc{Preview} triggered for instances where their message came across as sarcastic to \textit{"catch instances where you are trying not to sound sarcastic"} as \textit{"this definitely happened before"} with them. P7 and P8 noted that their inclination towards \textit{"lateral thinking"} and \textit{"jumping between topics"} sometimes led to others struggling to grasp the connection between their messages. They suggested \textsc{Preview} could proactively nudge users to stay focused on the topic at hand in such instances.

\paragraph{Tailoring \textsc{Preview} for Diverse Age Groups}
As a parent of autistic kids, P4 envisioned utilizing \textsc{Preview} to teach their children written communication skills. They contended that children of different age levels might require varying levels of explanations, \textit{"I think about how I talk to my 5-year-old autistic kid versus my 15-year-old [real age not disclosed to preserve anonymity] autistic kid, and how I explain things. People with different levels of experience will need different levels of instruction. So, if you have someone who has a lot of meta-cognitive, purpose-built social skills, or someone who doesn't have many of those skills, they might need different levels of explicit explanation. That might be something valuable to customize [with a slider]"}. In their opinion, more detailed feedback could entail precisely specifying the dependency between certain words/phrases and the recipient's reaction, \textit{"I would definitely need to explain to the 5-year-old why someone might perceive a message as blunt or dismissive or whatever. For example, my feedback could include `the message "horror is for kids" can be problematic because they [the receiver] don't want to be seen as a child, especially if they're an adult'. The extra explanation would be obnoxious to somebody who has already learnt that"}. P1 echoed with these suggestions. 

\paragraph{Personalizing Suggestions}
P2, P3, P4, and P8 stressed the importance of tailoring the suggested alternate messages to fit their individual writing styles, preferences, and interests. P4 suggested a simple strategy to enhance personalization, proposing that, in addition to offering new message suggestions, \textsc{Suggest} should also provide specific guidance on ways to improve the existing message. They saw this feature as a natural extension of \textsc{Suggest} seamlessly integrating with \textsc{Preview}. They elaborated, \textit{"it could provide insights like: You might be perceived as dismissive and judgmental of Ben's preferences -  A better message could be `I am not really into horror films', and then suggesting a genre that you do like."}. In their opinion, this could make room for personalization without the AI having to access their personal data, \textit{"how does the computer know what one likes without being connected to their data, which a lot of people might not be excited about?"}. In addition, P8 suggested having a calibration process \textit{"[for the application to] understand me a little bit more"} by \textit{"figuring out what my tone normally is"}. 

\section{Discussion}
\label{sec:discussion}
In this section, we delve into the design and practical implications of our study, explore future directions, and discuss our limitations.

\subsection{Balancing Personalization and Privacy}

Participants highlighted the need for improved personalization by adjusting flagging sensitivity, tailoring feedback to users' abilities, and matching suggestions to their writing styles, alongside concerns that personalization might require extensive collection of user data. It is important to note that different levels of personalization require varying degrees of user data. Adapting to a user's writing style in a certain conversation involves analyzing how they communicate within that conversation \cite{ghostwriter-yeh-catherine}. Conversely, providing highly personalized message suggestions based on specific interests requires the system to know about users beyond the content of a single conversation \cite{roy-a-systematic}. Hence, it is crucial to ask a) with access to `just enough information' about users, can we achieve the required level of personalization, and b) what constitutes `just enough information'. In the context of \textsc{Suggest}, one strategy is to provide `guidance' or `advice' rather than suggesting alternative messages as replacements. This allows users to personalize their message as well as receive help on how to write and what kind of content to include (or not), without needing to reveal personal data beyond the immediate context of the conversation.  Another option could be to iteratively obtain user feedback and regenerate the suggestions accordingly. Personalization is crucial, yet our findings also indicate that giving users too much to manage—such as overly frequent underlining of phrases—can be overwhelming. Therefore, a key aspect of inculcating personalization into systems lies in offering users control over customizable knobs without over-burdening them \cite{hui-enhancing-user}.

\subsection{Toward Trustworthy User Interfaces}
Recall that participants were hesitant to trust the system too much, knowing it may not always be correct. They feared that using the system for longer periods could lead them to rely on it even in instances where it was wrong. In the context of providing social support, particularly through subjective interpretations and judgments, being strictly right or wrong is challenging as the connection between writing style and intent can be unclear depending on context and individual differences \cite{thomas-text-messaging}. Our findings underscore the importance of clearly and transparently communicating this uncertainty to the user, which is inherent to judgements of this nature, particularly when AI is making the judgments. Effectively utilizing user interfaces might be one way to address this \cite{chen-is-this}. Visual indicators of the AI's confidence in its output, or using language that suggests possibility (`could', `might') rather than certainty (`will', `must'), can remind users that the AI may not always be correct and encourage them to trust their own instincts. In addition, incorporating adaptive learning in TwIPS by tailoring its assistance to individual user needs could promote independence and decrease users’ reliance on it over time. As users gain experience, they might need less assistance, while new users might need more.

\subsection{Beyond Neurodivergent Users}
Drawing inspiration from the double empathy problem \cite{damian-on-the}, we advocate for measures that encourage neurotypical individuals to also contribute to improving communication. P4 and P7 expressed a strong desire for such measures too. For P4, it would be \textit{“interesting to see an app that goes the other way around - rewriting messages for neurotypicals to incorporate direct language.”} P7 echoed these sentiments, expressing, \textit{"They [neurotypical individuals] sort of need to meet me halfway!”} Such an approach could involve deploying a version of the TwIPS prototype on \textit{both} ends of chatting applications for \textit{all} users, whether they identified as neurodivergent or not. An autistic user could receive help from this prototype with interpreting nuanced language and emojis. The same prototype could help a neurotypical user understand the unique writing style of their autistic peer, explaining that a brief reply does not necessarily signal disinterest. This could also extend to interactions among neurotypical users, considering emotional expression and interpretation in text messages, although not to the same degree as autistic users, is challenging for a large proportion of all users \cite{miller-understanding-emoji, lynne-perceived-miscommunication}. In this way, the prototype could provide feedback to \textit{any} user based on the specific communication challenges they face and the writing style of the person they are interacting with. Realizing this vision would require accessibility researchers to engage with neurotypical individuals, exploring their perceptions and interactions with neurodivergent individuals, and evaluate prototypes collectively with all users involved. \cite{morris-double-empathy}. Given communication is fundamentally collaborative, it would be interesting for researchers to explore how double empathy can serve as a design framework to extend this approach to other modes of digital communication. 



\subsection{Combining AI-based Simulations with Real-world Interactions}
TwIPS acted as both a learning tool and a safeguard. For example, participants expressed \textsc{Preview} not only helped them develop a mental model of how their words might be perceived as rude, dismissive, or offensive by others (like a learning tool), but also prevented them from coming across as such (like a safeguard). AI-aided simulations together with AI-assistance - similar to participants' interaction with Ben's AI-persona in phase 2 -  can provide an environment to learn, and real-time AI-assistance during real-world interactions can serve as a safeguard. 
However, combining AI-aided simulations with real-world interactions could create a more long-lasting, personalized and dynamic tool. Different AI-personas could allow autistic individuals to engage with different kinds of conversational partners and scenarios in a controlled environment, akin to past AR-based simulations that exposed users to different physical environments or social stories \cite{exploring-design-ahsen}. Conversely, assistance in real-world interactions could help users mitigate challenges in situations involving complex and varied stakes, and enhance learning in real-world settings which may differ from simulations. Together, the simulations and real-world interactions form a symbiotic relationship, where real-world interactions help inform the design of specific simulation strategies while simulations inform the level of assistance needed and what to focus on during real-world interactions. It would be useful for future researchers to explore how to best leverage this symbiotic relationship within a single, autonomous system.

\subsection{Practical Implications and Lessons Learnt}
Participants' reactions to our prototype suggest that TwIPS could add value to texting platforms, dating applications, and social media sites that support text-based communication. Popular platforms like Meta's WhatsApp already support generative AI-based bots for open-ended QnA \cite{company-introducing-new} and possess the technical infrastructure needed for widespread AI-adoption. This existing infrastructure, combined with the fact that TwIPS was designed with the layout of standard chatting applications in mind, makes them well-suited for incorporating TwIPS' features. However, any integration would need to be done in a privacy-preserving manner \cite{yiming-privatelora-for}. While there are benefits to deploying it on scale ourselves, such as control over functionality and design, and the opportunity to conduct longitudinal user studies with more participants, practical constraints such as cost pose a significant feasibility challenge for academic researchers.
Cost to run the LLM was a major contributor to our expenses, which varies with the length of the input and output of each LLM call as well as model quality, which depends on model size \cite{jared-scaling-laws} and training data quality \cite{tianxing-from-data}. We fed the complete conversation history to GPT4 with each call, causing the input length to increase substantially. Employing a cheaper model \cite{touvron-llama-2} or retrieval augmented generation techniques \cite{patrick-retrieval-augmented} to extract relevant portions of history could shorten the input length and thus lower cost. In addition, we observed that one LLM does not need to perform all tasks. Given the variety of models available with different costs and quality, strategically allocating advanced, expensive models for complex tasks, and choosing affordable, less sophisticated models for other use-cases can significantly reduce cost. In the context of TwIPS, a complex task might involve using \textsc{Preview} for a conversation with a potential date as opposed to a casual chat with a friend, or utilizing \textsc{Interpret} for explaining a hyper-local phrase unique to a conversation rather than a well-known idiom.

\subsection{Limitations}

There are a number of limitations of our study. Recruiting participants solely from a university setting restricts the generalizability of our findings to the broader autistic community, due to a lack of diversity in age, background, and education. While the data from our participant pool showed repeated themes, expanding it to include a more diverse demographic might uncover additional themes. In both phases of the user study, participants engaged with an imaginary character. Participants may have made varying assumptions about their relationship dynamics with Ben, leading to differences in their interactions and responses. While keeping the complexity of this setup minimal, we made maximum effort to provide all essential details of the setup to participants. Lastly, the in-lab setting of our study may not fully replicate the nuances and dynamics of real-life texting, which typically involves more complex and varied stakes. A longer-term deployment and evaluation of TwIPS in the wild, paired with a control group comparison, could yield a more confident assessment.

\section{Conclusion}

In this paper, we presented the design and evaluation of TwIPS, a prototype texting application powered by a large language model to simplify conversational nuances for autistic users. We evaluated our prototype with 8 autistic participants in an in-lab setting. Our findings revealed that TwIPS enabled a convenient way for participants to seek clarifications, provided a better alternative to tone indicators, and facilitated constructive reflection on writing technique and style. We also examined how
autistic users utilize language for self-expression and interpretation in instant messaging, and gathered feedback for enhancing our prototype. We concluded with a discussion around balancing user trust and autonomy with AI-assistance, users’ customization needs in the context
of AI-assisted communication, and designing interventions that distribute responsibility for reducing communication breakdowns more equitably across all users, instead of placing it solely on autistic users.


\bibliographystyle{unsrt}
\section{Acknowledgments}

We are grateful to our participants, the ASSETS reviewers, and all members of NAT Lab, Crehan Lab, and D.O.C.C. Lab at Tufts University for their feedback on this work. We are also thankful to Dr. Elaine Schaertl Short for her early guidance and sharing her perspective on disability and autism research. This work was partially supported by a Tufts Springboard award.

\clearpage
\appendix
\onecolumn
\section{Prompt Templates and Flows}
\label{sec:appendix-a}
\begin{figure*}[htbp]
    \centering
    \includegraphics[width=0.9\textwidth]{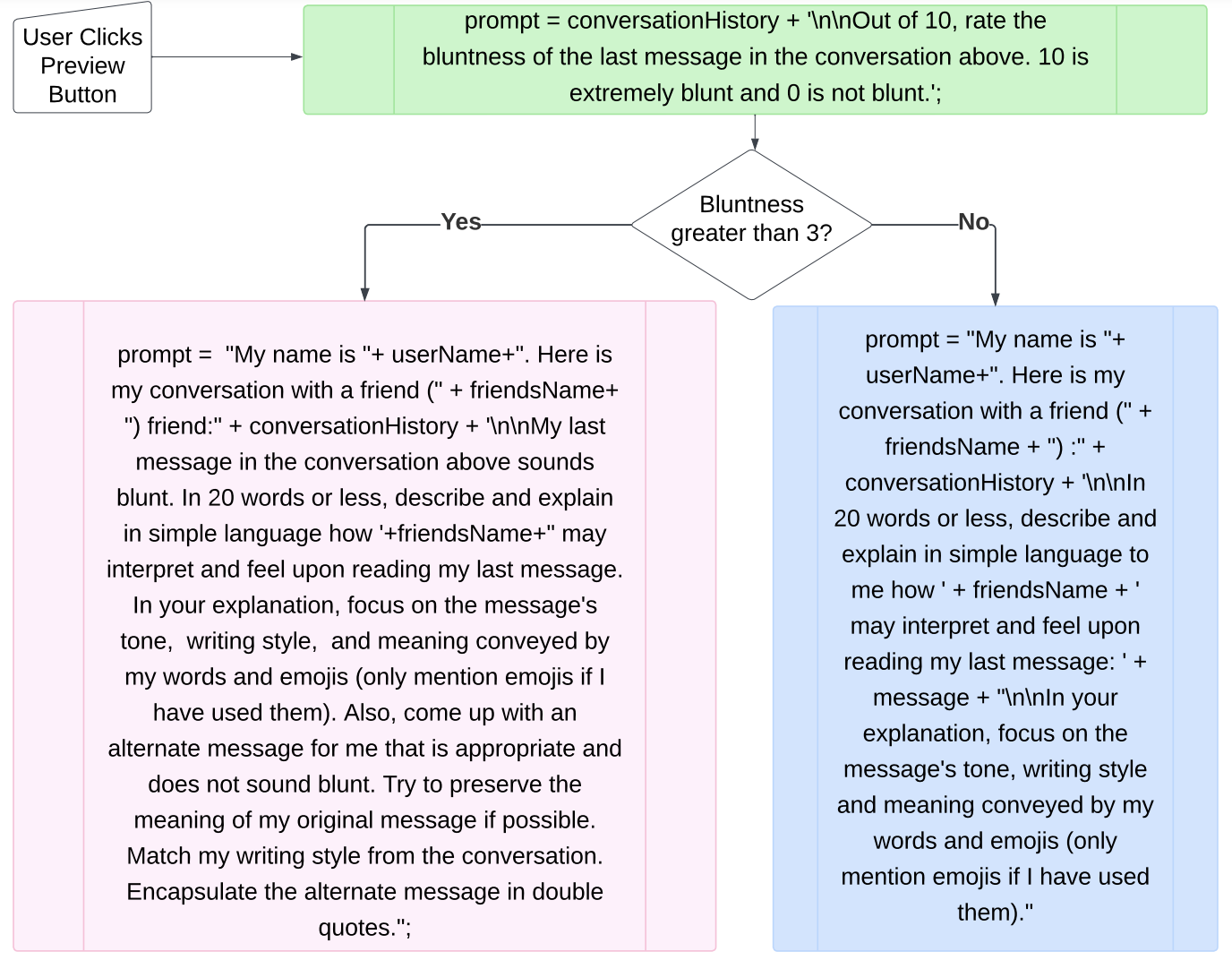}
    \Description{This figure shows the prompt template and flow for clicking on the Preview button. When the user clicks on the preview button, an LLM call is made to determine the bluntness of the user's message out of 10. If the bluntness is greater than 3, another LLM call is made to generate a preview of the user's likely reaction, as well as a less blunt, alternate suggestion. If it is less than 3, only the preview is generated in a separate call.}
    \caption{Prompt template and flow for clicking on `Preview Button'.}
    \label{fig:preview-flow}
\end{figure*}

\begin{figure*}[htbp]
    \centering
    \includegraphics[width=0.8\textwidth]{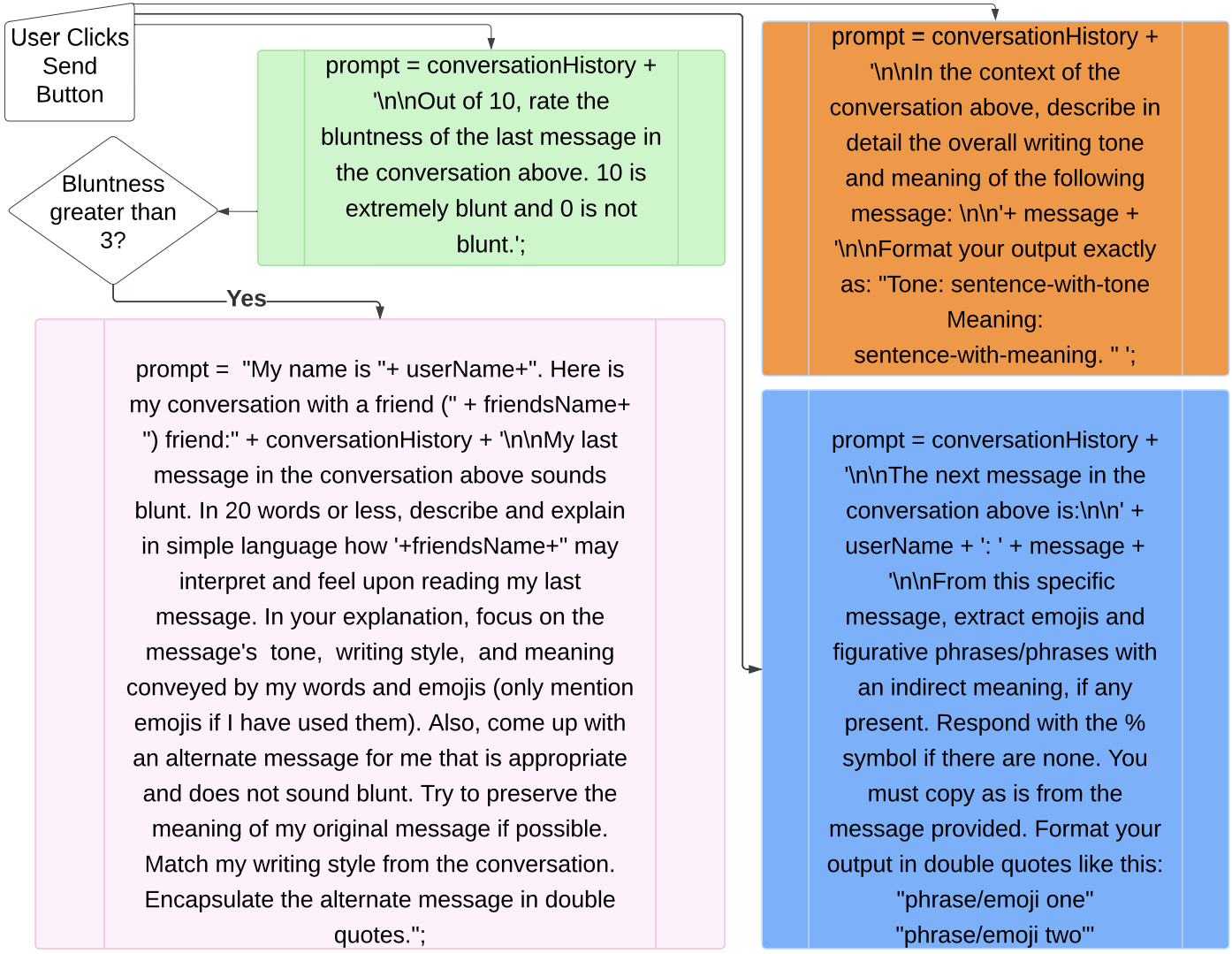}
    \Description{
    This figure shows the prompt template and flow for clicking on the Send button. When the user clicks on the preview button, an LLM call is made to determine the bluntness of the user's message out of 10. If the bluntness is greater than 3, another LLM call is made to generate a preview of the user's likely reaction, as well as a less blunt, alternate suggestion. If not, no further processing happens. At the same time, a call to the generate the overall tone and meaning of the message and another call to identify any ambiguous language element is made.}
    \caption{Prompt template and flow for clicking on `Send Button'.}
    \label{fig:send-flow}
\end{figure*}

\begin{figure*}[htbp]
    \centering
\includegraphics[width=0.6\textwidth]{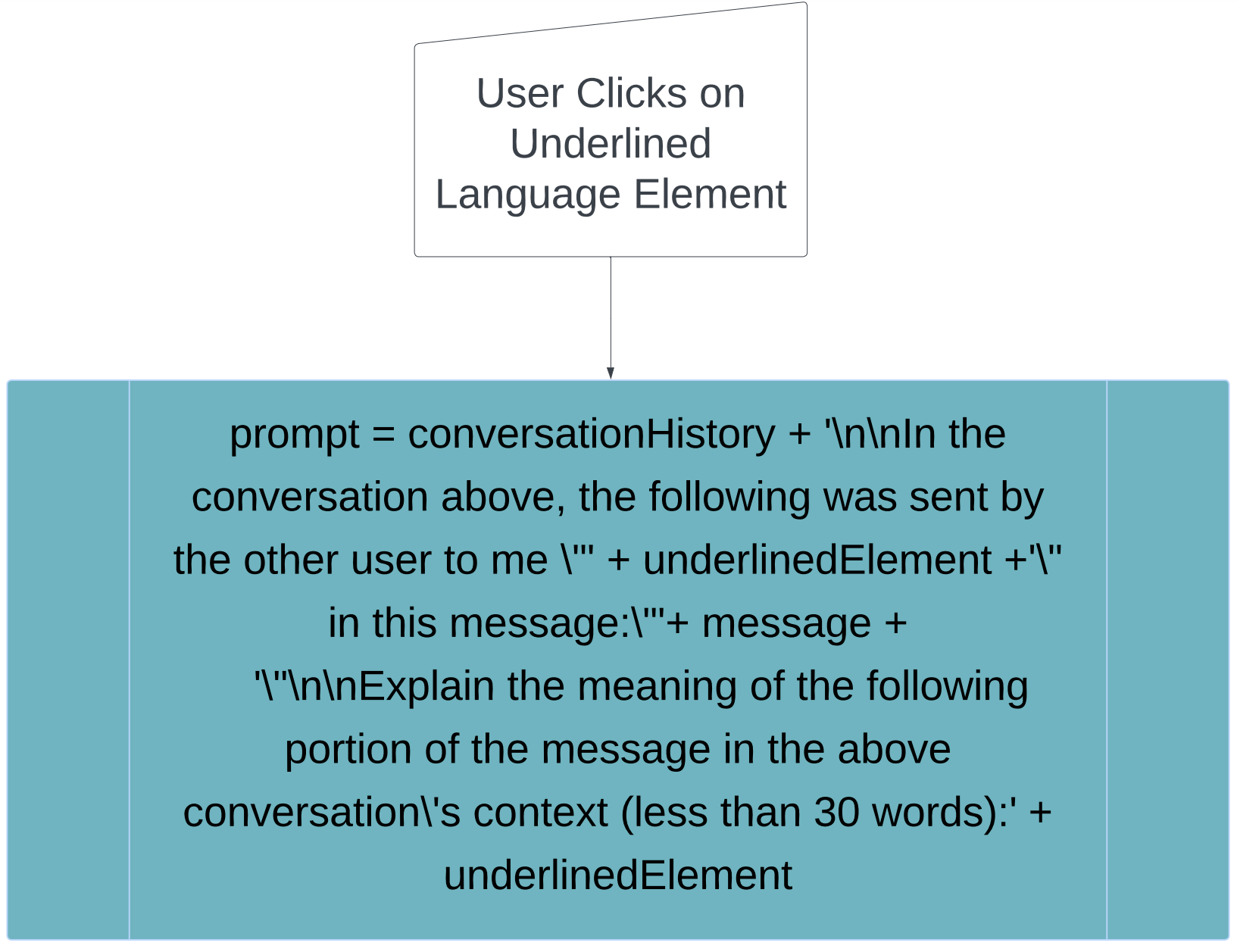}
\Description{This figure shows the prompt template and flow for clicking on an underlined language element in a message. Upon clicking, an LLM call to explain that language element is made.}
    \caption{Prompt template and flow for clicking on an underlined ambiguous language element.}
    \label{fig:interpret-flow}
\end{figure*}

Figures \ref{fig:preview-flow}, \ref{fig:send-flow} and \ref{fig:interpret-flow} show the prompt templates and flows used within TwIPS.

\clearpage

\section{Conversational Script Used in Phase 1}
\label{sec:appendix-b}
The conversational script used in phase 1 of the user study is provided below.
\vspace{0.5cm}

\setspeaker{Ben}
\setspeaker{Participant}[Model Response]
\addtolength{\transcriptlen}{1em}%

\begin{description}

  \Ben hey, did you hear? it's Jack's birthday next week!
  \Participant yeap, i know!

  \Ben well, I thought we could plan a surprise party for him. 
  \Participant hmm.. is a party our only option?

  \Ben nope, we can do something else too. like a movie night at the theater.
  \Participant we could do that, yes!
  
  \Ben awesome! we can pick a movie that everyone likes. Are you into horror films?
  \Participant no... horror is for kids who want to be scared
  
  \Ben LOL. sometimes a good horror flick can be a fun experience, even for grown ups.
  \Participant if it will make jack happy, i am ok with it i guess
    
  \Ben by the way, have you thought about what kind of birthday present we should get him?
  \Participant can we just get him a gift card?
  
  \Ben it might be nice to show him that we put some thought into his bday present. it's his birthday, and he's our friend!!
  \Participant yeah, you are actually right. we can go to the store tomorrow to see what we can get.
  
  \Ben we can split costs. what is your budget?
  \Participant I do not want to spend too much.
  
  \Ben thats fine... There is a store at a 20 min walk from the office. We can go there.
  \Participant alright, I'll put on my joggers then haha
  
  \Ben LOL! Should we invite others too?
  \Participant umm... is that necessary?
  
  \Ben nope, not really
  \Participant so let’s not, I guess.
  
  \Ben why though?
  \Participant coordinating with others is a hassle
  
  \Ben Lol okay if you say so! see you in class tomorrow then.
  \Participant  oki! catch you later :)
\balance
\end{description}
\balance

\end{document}